%% file: paper.tex
\documentclass{llncs}
\usepackage{amsmath,amssymb,mathrsfs}            
\usepackage{graphicx}                            

\usepackage[linesnumbered,ruled,vlined]{algorithm2e}
\DontPrintSemicolon%
\SetKwInOut{Input}{in}
\SetKwInOut{Output}{out}
\SetKwFunction{solve}{solve}
\SetKwFunction{atleast}{atleast1}
\SetKwFunction{choice}{choice}
\SetKwFunction{pick}{pick}
\SetKwFunction{condition}{cond}
\SetKwData{res}{res}
\SetKwData{truec}{true}
\SetKwData{nullc}{null}

\newtheorem{observation}{Observation}

\def\miko{Mikol\'a\v{s} Janota}
\def\jpms{Joao Marques-Silva}
\def\theTitle{On Minimal Corrections in ASP}
\def\theTitleBroken{On Minimal Corrections in ASP}
\DeclareMathOperator*{\aspnot}{\textit{not}\hspace{1pt}}
\DeclareMathOperator*{\grd}{\textit{grd}}

\DeclareMathOperator*{\cS}{\mathcal{S}}
\DeclareMathOperator*{\cL}{\mathcal{L}}
\DeclareMathOperator*{\cA}{\mathcal{A}}
\DeclareMathOperator*{\cR}{\mathcal{R}}
\DeclareMathOperator*{\cC}{\mathcal{C}}
\DeclareMathOperator{\emptyPred}{\textit{empty}}
\DeclareMathOperator{\fullPred}{\textit{full}}
\DeclareMathOperator{\locationPred}{\textit{location}}
\DeclareMathOperator{\move}{\textit{move}}
\DeclareMathOperator{\stone}{\textit{stone}}

\DeclareMathOperator{\head}{\textit{head}}
\DeclareMathOperator{\body}{\textit{body}}
\DeclareMathOperator{\union}{\cup}
\def\rr{\mathrm{r}}
\def\ra{\mathrm{a}}
\def\problem{\textsc{MinCorrect}\xspace}
\newcommand{\comprehension}[2]{\ensuremath{\left\{ {#1} \;|\; {#2}\right\}}}
\usepackage[pdftex]{color}
\definecolor{citeblue}{rgb}{0.1,0,.4}
\definecolor{refcolor}{rgb}{0,0,0.4}
\definecolor{midgreen}{RGB}{0,150,0}
\definecolor{darkgreen}{RGB}{0,128,0}
\definecolor{darkblue}{RGB}{0,0,128}
\definecolor{darkred}{RGB}{192,0,0}
\usepackage[pdftex%
,colorlinks=true%
,bookmarks=true%
,linkcolor=citeblue%
,citecolor=citeblue%
,urlcolor=blue%
,plainpages=false]{hyperref}
\hypersetup{%
pdfauthor={{\miko \jpms}},
pdftitle={{\theTitle}} }

\title{\theTitleBroken}
\author{
{\miko} \inst{1}\!
\and {\jpms} \inst{1,2}
}

\institute{
IST/INESC-ID, Lisbon, Portugal
\and
University College Dublin, Ireland}

\begin{document}
\maketitle
\begin {abstract}
\input{abs}
\end {abstract}
\input{introduction}

\input{preliminaries}

\input{mmodels}

\input{mcs}
\input{experiments}
\input{related}
\input{conc}
\section*{Acknowledgment}
This work is partially supported by SFI grant BEACON
(09/IN.1/\-I2618), by FCT grant POLARIS (PTDC/EIA-CCO/123051/2010),
and by INESC-ID's multiannual PIDDAC funding PEst-OE/EEI/LA0021/2013.

\end{document}

%% file: abs.tex

As a programming paradigm, answer set programming (ASP) brings about the usual
issue of the human error. Hence, it is desirable to provide automated techniques
that could help the programmer to find the error.
This paper addresses  the question of computing a subset-minimal correction of
a contradictory ASP program. A contradictory ASP program is often undesirable and
we wish to provide an automated way of fixing it.
We consider a minimal correction set of a contradictory
program to be an irreducible  set of rules whose removal makes the program
consistent. In contrast to propositional logic, corrections of ASP programs
behave non-monotonically. Nevertheless, we show that a variety of algorithms
for correction set computation in propositional logic can be ported to ASP.
An experimental evaluation was carried showing that having a portfolio of
such algorithms is indeed of benefit.

%% file: introduction.tex
\section{Introduction}
 Answer set programming (ASP) is a powerful paradigm for modeling and solving
 combinatorial and optimization problems in artificial intelligence. An inconsistent program is such a program that does not have a solution.
 This might be due to a bug in the program and for such cases it is desirable to
 provide the programmer with tools that would help him to identify the issue.

 The primary motivation for this paper is to focus on program's \emph{input}.
 In theory, an input to an ASP program is really just another ASP program that is
 joined to the original one. In practice, however, the conceptual division between a
 program and its input plays an important role in the program's development.  Indeed, a
 program, as a set of rules, expresses the semantics of the problem being solved. The
 input is a set of \emph{facts} describing a particular instance of the problem. Such input is
 typically large.

 This paper asks the question, how to resolve situations when there is an error in the given input? In particular, we
 consider scenarios when the input leads to a contradiction in the program.
 Consider the following simple program.
 \begin{align}
   & \leftarrow \aspnot\move(a).&\texttt{\% program}\label{eq:g}\\
   \move(a)&\leftarrow\stone(b),\aspnot\stone(c).&\texttt{\% program}\label{eq:mv}\\
   \stone(c) & \leftarrow .&\texttt{\% input}\label{eq:sc}
 \end{align}
 By rule~\eqref{eq:g} the program requires $\move(a)$ be true. To achieve that, however,
 $\stone(b)$ must be true and $\stone(c)$ must be false (by rule~\eqref{eq:mv}).
 The input is specified as the fact~\eqref{eq:sc}. Altogether, the program is
 \emph{inconsistent} since rule~\eqref{eq:mv} is not applicable. If we wish to modify the input so it becomes
 consistent, the fact $\stone(c)$ must be removed and the fact $\stone(b)$ must be added.
 It is the aim of this paper to compute such corrections to inconsistent inputs.
 Further, we aim at corrections that are \emph{irreducible}, i.e.\ that do not perform unnecessary changes.
 Note that this program could also be made consistent by modifying the rules \eqref{eq:g}
 and \eqref{eq:mv}. Similarly, one could add the fact $\move(a)$. This might
 be undesirable as these represent the rules of the modeled move (in a game, for
 instance). In the end, however, it is the programmer that must decide what  to
 correct. The objective of the proposed tool-support is to pin-point the source
 of the inconsistency.

 We show that the problem of inconsistency corrections is closely related to a
 problem of \emph{maximal consistency}, which we define as identifying a
 subset-maximal set of atoms that can be added to the program as facts while
 preserving consistency.  Maximal consistency lets us provide a solution to
 inconsistency correction but it is also an interesting problem to study on its on.

 Maximal consistency is closely related to the concept of \emph{maximally satisfiable sets (MSS)} in propositional logic.
 For a formula in conjunctive normal form, an MSS is a subset of the formula's clauses that is satisfiable and adding any clause
 to it makes it unsatisfiable.  There is also a dual, minimal correction subset (MCS), which is a complement of an MSS~\cite{liffiton-jar08}.

 There is an important difference between propositional logic and ASP
 and that is that ASP is \emph{not monotone}.  This means that an
 algorithm for calculating MSSes cannot be immediately used for ASP.
 We show, however, that it is possible to port existing MSS
 algorithms to ASP.  This represents a great potential for calculating
 maximally consistent sets in ASP as a bevy of algorithms for MSS
 exist~\cite{DBLP:conf/padl/BaileyS05,mshjpb-ijcai13,msjb-cav13}.

 The main contributions of this paper are the following.
 (1)~It devises a technique for adapting algorithms from MSS computation to maximal consistency in ASP.
 (2)~Using the technique a handful of MSS algorithms is adapted to ASP.
 (3)~It is shown how maximal consistency can be used to calculate minimal
 corrections to ASP inputs.
 (4)~The proposed algorithms were implemented and evaluated on a number of benchmarks.

%


The paper is organized as follows.
\autoref{sec:mmodels} introduces the concept of maximal consistent subsets and
proposes a handful of algorithms for computing them.
\autoref{sec:mcs} relates maximal consistent subsets to corrections of
programs.
\autoref{sec:experiments} presents experimental evaluation of the presented
algorithms.
\autoref{sec:related} briefly overview related work--in ASP but also in
propositional logic.
Finally, \autoref{sec:conc} concludes and discusses topics of future
work.

%% file: preliminaries.tex
\section{Background}

We assume the reader's familiarity with standard ASP syntax and semantics, e.g.~\cite{baral2003knowledge}.
Here we briefly review the basic notation and concepts.
In particular, a \emph{(normal) logic program} is a finite set of \emph{rules} of the
following form.
\[a\leftarrow b_1,\dots,b_m,\aspnot c_{m+1},\dots,\aspnot c_{n},\]
where $a,b_i,c_j$ are atoms.
A \emph{literal} is an atom or its default negation $\aspnot a$.
A rule is called a \emph{fact} if it has an empty body; in such case we don't write the symbol~$\leftarrow$.
For a rule~$r$, we write \emph{$\body(r)$} to denote the literals~$b_1,\dots,b_m,\aspnot c_{m+1},\dots,\aspnot c_{n}$
and we write \emph{$\head(r)$} to denote the literal~$a$.
We write $B^+(r)$ for $b_1\dots,_m$ and $B^-(r)$ for $c_{m+1},\dots,c_n$.
Further, we allow \emph{choice rules} of the form \[n\leq\{a_1,\dots,a_k\}.\]
(this is a special case of \emph{weight constraint rules}~\cite{DBLP:conf/lpnmr/Simons99,DBLP:conf/lpnmr/NiemelaSS99,DBLP:journals/ai/SimonsNS02}).

A program is called \emph{ground} if does not contain any variables.
A \emph{ground instance} of a program~$P$, denoted as~$\grd(P)$,
is a ground program obtained by substituting variables of $P$ by all constants from its Herbrand universe.

The semantics of ASP programs can be defined via a \emph{reduct}~\cite{DBLP:conf/iclp/GelfondL88,DBLP:journals/ngc/GelfondL91}.
Let $I$ be a set of ground atoms.
The set $I$ is a \emph{model} of a program $P$ if
$\head(r)\in I$ whenever $B^+(r)\subseteq I$ and $B^-(r)\cap I=\emptyset$
for every $r\in\grd(P)$.
The reduct of a program~$P$ w.r.t.\ the set~$I$ is denoted as $P^I$
and defined as follows. 
  \[P^I=\comprehension{\head(r)}{r\in\grd(P), I\cap B^-(r)=\emptyset}\]

The set~$I$ is an \emph{answer set} of~$P$ if~$I$ is a minimal model of~$P^I$.
This definition guarantees that an answer set contains only atoms that have an
acyclic justification by the rules of~$P$ (cf.~\cite{DBLP:conf/ijcai/Lee05}).
A choice rule $l\leq\{s_1,\dots,s_k\}.$ in a program additionally guarantees that
any answer set contains at least $l$ atoms from $s_1,\dots,s_k$,
and, the rule provides a justification for any of those atoms.
For precise semantics see~\cite{DBLP:conf/lpnmr/NiemelaSS99,DBLP:journals/ai/SimonsNS02}.
A program is \emph{consistent} if it has at least one answer set,
it is \emph{inconsistent} otherwise.

%


%% file: mmodels.tex
\section{Maximal Consistency in ASP}\label{sec:mmodels}
This section studies the problem of computing a maximal subset of given atoms
whose addition to the program, as facts, yields a consistent program.
\begin{definition}[maximal consistent subset]
Let~$P$ be a consistent ASP program and~$\cS$ be a set of atoms.
A set~$\cL\subseteq\cS$ is a \emph{maximal consistent subset} of $\cS$ w.r.t.\ $P$
if the program $P\union\comprehension{s.}{s\in\cL}$
is consistent and for any $\cL'$, such that $\cL\subsetneq\cL'\subseteq\cS$,
the program $P\union\comprehension{s.}{s\in\cL'}$ is inconsistent.
\end{definition}

The definition of maximal consistent subset is syntactically similar to
\emph{maximal models} or MSSes in propositional logic. Semantically, however, there is an
important difference due to nonmonotonicity of ASP.  While in propositional
logic the satisfiability of $\phi\land x\land y$ guarantees satisfiability of
$\phi\land x$, in ASP it is not necessarily the case. Hence, algorithms for
MSSes in propositional logic \emph{cannot} be readily used for our maximal consistent subset.
We will show, however, that it is possible to port these algorithms to ASP.

\input{algs}

%% file: algs.tex
In the following we use some auxiliary functions.
The function $\choice(\cS)$ produces the choice rule
$0\leq\{s_1,\dots,s_k\}$ over the set $\cS=\{s_1,\dots,s_k\}$.
The function~$\atleast(\{s_1,\dots,s_k\})$ produces the choice rule
$1\leq\{s_1,\dots,s_k\}$. 
An ASP solver is modeled by the function~$\solve(P)$, which
returns a pair~$(\res,\mu)$ where~$\res$ is \textit{true} if and only if~$P$
is consistent and~$\mu$ is an answer set of~$P$ if some exists.

A \emph{brute force} approach to calculating a maximal consistent subset would be to enumerate all subsets of $\cS$
and for each test whether it is still consistent. As there are $2^{|\cS|}$ subsets of $\cS$,
this approach is clearly unfeasible.

A better approach is to \emph{maximize} the sum~$\Sigma_{s\in S} s$ with respect to the program~$P$.
Such ensures finding a maximal consistent subset with \emph{maximum cardinality}.
This can be done by iteratively calling an ASP solver while imposing increasing cardinality on the set $\cS$.
However, modern ASP solvers directly support minimization constraints through which
maximization can be specified by minimizing negation of the atoms in $\cS$.
This approach was taken elsewhere~\cite{syrjanen2006debugging,DBLP:conf/aaai/GebserPST08}.

Finding a maximal correction subset with maximum cardinality might be computationally harder
than finding \emph{some} maximal correction subset. This is the purposed of the rest of the section.

We begin by an important observation that it is possible to check whether a set of atoms~$\cL$ can be extended into a
consistent set of atoms by a single call to an ASP solver.

\begin{observation}\label{obs:extension}
  Let~$P$ be an ASP program and $\cL\subseteq\cS$ be sets of atoms.
  Let~$P'$ be defined as follows.
   \[P' = P\union\comprehension{s.}{s\in\cL}\union\{\choice(\cS\smallsetminus\cL).\}\]
  There exists a set of atoms $\cL'$ s.t.\ $\cL\subseteq\cL'\subseteq\cS$
  iff $P'$ has an answer set~$\mu$ such that~$\cL'=\cS\cap\mu$.
\end{observation}

\autoref{obs:extension} enables us to check whether a set of atoms can be extended into a consistent set.
This is done by letting the solver to chose the additional atoms.
Similar reasoning enables us to devise a test for checking that a consistent set~$\cL$ is already maximal.
This is done by enforcing that~$\cL$ is extended by at least one element.

\begin{observation}\label{obs:termination}
  Let $P$ be an ASP program and $\cS$ be a set of atoms.
  Let $P'$ be defined as follows.
  \[P' = P\union\comprehension{s.}{s\in\cL}\union\{\atleast(\cS\smallsetminus\cL)\}\]
  A set $\cL\subseteq\cS$ is a maximal consistent subset of $\cS$
  iff $P\union\comprehension{s.}{s\in\cL}$ is consistent and $P'$ is inconsistent.
\end{observation}

\input{alg_a}%

Combining \autoref{obs:extension} and \autoref{obs:termination} gives us \autoref{alg:a}.  The algorithm maintains a
lower bound $\cL$, which is initialized to the empty set.  The set $\cL$ grows incrementally until a maximal set is
found.  In each iteration it is tested whether~$\cL$ is already maximal in accord with \autoref{obs:termination}.
The invariant of the loop is that the program $P\union\comprehension{s.}{s\in\cL}$ is consistent.  The set $\cL$
grows monotonically with the iterations because whenever a new answer set $\mu$ is obtained, it must contain all the
atoms from the previous $\cL$. Hence, the algorithm terminates in at most $|S|$ calls to an ASP solver. Note that the
algorithm does not necessarily need all of the~$|S|$ iterations since more than one element might be added to $\cL$
in one iteration.
Similar algorithms have been proposed in different contexts, including
computing the backbone of a propositional
formula~\cite{malik-hldvt11,malik-fmcad11} and computing an
\emph{minimal correction set (MCS)} of a propositional
formula~\cite{mshjpb-ijcai13}.

\input{alg_u}
The second algorithm we propose also maintains a lower bound on a maximal
consistent set and tests, one by one, the elements to be added. The pseudo-code can be
found in \autoref{alg:u}. In each iteration it tests whether an
element~$s_f$ can be added to the current lower bound~$\cL$. If it is possible,
the algorithm continues with a larger $\cL$. If it not, then $s_f$ is removed
from $\cL$ and never inspected again.

In contrast to \autoref{alg:a}, here it is
not immediately clear that the returned set is indeed maximal. Indeed, how do we
know that if it was impossible to add  some element~$s_f$ to an earlier~$\cL_1$,
that is it still impossible for the final~$\cL_f$? This follows from
\autoref{obs:extension} and from the fact that $\cL$ grows monotonically
throughout the algorithm. More precisely, it holds that $\cL_1\subseteq\cL_f$
and our test checks that $s_f$ cannot be added to any \emph{superset} of $\cL_1$,
i.e.\ if it was \emph{not} possible to extend $\cL_1$ with $s_f$, then it is also
impossible to extend $\cL_f$ with it. As \autoref{alg:a}, \autoref{alg:u}
also performs at most $|\cS|$ calls to an ASP solver.
Similar algorithms have been proposed in a number of contexts,
including computing an MCS~\cite{DBLP:conf/padl/BaileyS05,biere-vamos12}.

The third algorithm we consider is inspired by the \emph{progression
  algorithm}~\cite{msjb-cav13}.
The algorithm is shown in \autoref{alg:p}.
It tries to add more than a single atom to the lower bound in one iteration---atoms are
added in \emph{chunks}.  This is done progressively: in the very first iteration the chunk contains
a single atom. The chunk size is  doubled each time the current chunk is added with success.
If it is not possible to extend the current lower bound $\cL$ with the current chunk,
the size is reset again to $1$. Whenever a chunk of size $1$ cannot be added to $\cL$
the clause comprising the chunk is no longer inspected. This guarantees termination.
Note that the algorithm aims at constructing $\cL$ as quickly as possible by
adding larger chunks of atoms at a time.
%
\input{alg_p}
The use of chunks finds other applications, including redundancy
removal~\cite{bjlms-cp12}.

%% file: alg_a.tex
\begin{algorithm}[t]
$\cL\gets\emptyset$\;
\While{\truec} {
  $P'\gets P\cup\comprehension{s.}{s\in\cL}$\;
  $P'\gets P'\cup\{\atleast(\cS\smallsetminus\cL).\}$\;
  $(\res,\mu)\gets\solve(P')$\;
  \lIf{$\lnot \res$}{\Return $\cL$}
  $\cL\gets\mu\cap\cS$\;
}
\caption{Atleast-1 algorithm for maximal consistency.}\label{alg:a}
\end{algorithm}

%% file: alg_u.tex
\begin{algorithm}[t]
$\cL\gets\emptyset$\;
\While{$\cS\neq\emptyset$}{
  $s_f\gets\text{ pick an arbitrary element from }\cS$\;
  $\cS\gets\cS\smallsetminus\{s_f\}$\;
  $\cL\gets\cL\union\{s_f\}$\;
  $P'\gets P\union\{ \choice(\cS). \} $\;
  $P'\gets P'\union\comprehension{s.}{s\in\cL}$\;
  $(\res,\mu)\gets\solve(P')$\;
  \lIf{$\lnot\res$}{$\cL\gets\cL\smallsetminus\{s_f\}$}
  \lElse{$\cL\gets\mu\cap\cS$}
}
\Return $\cL$
\caption{Iterative algorithm for maximal consistency.}\label{alg:u}
\end{algorithm}

%% file: alg_p.tex
\begin{algorithm}[t]
$\cL\gets\emptyset$\tcp*[r]{consistency lower bound}
$K\gets 1$\tcp*[r]{chunk size}
\While{$\cS\neq\emptyset$}{
  $\cC\gets\text{pick }\textrm{min}(|\cS|,K)\text{ arbitrary elements from }\cS$\;
  $\cS\gets\cS\smallsetminus\cC$\;
  $\cL\gets\cL\union\cC$\;
  $P'\gets P\union\{ \choice(\cS). \} $\;
  $P'\gets P'\union\comprehension{s.}{s\in\cL}$\;
  $(\res,\mu)\gets\solve(P')$\;
  \If{$\lnot\res$}{\tcp*[l]{re-analyze chunk more finely}
    $\cL\gets\cL\smallsetminus\cC$\;
    \lIf{$K>1$}{$\cS\gets\cS\union\cC$}
    $K=1$\tcp*[r]{reset chunk size}
  } \Else{
    $K\gets 2K$\tcp*[r]{double chunk size}
    $\cL\gets\mu\cap\cS$\tcp*[r]{observe that $\cC\subseteq(\mu\cap\cS)$}
  }
}
\Return $\cL$
\caption{Progression-based algorithm for maximal consistency.}\label{alg:p}
\end{algorithm}

%% file: mcs.tex
\section{From Maximal Consistency to Minimal Corrections}\label{sec:mcs}
 In this section we see how maximal consistency is useful for calculating
 corrections to an inconsistent program. Hence, the objective is to calculate an
 irreducible correction to a given inconsistent program so it becomes
 consistent.

 The concept of \emph{(minimal) correction sets} commonly appears in
 propositional logic~\cite{DBLP:conf/padl/BaileyS05}.  Correction
 sets, however, are meant to be removed from the formula in order to make it
 consistent.  In ASP, however, corrections might consist of removal or \emph{addition}.
 Indeed, unlike in propositional logic, an ASP program may become consistent after a fact (or rule) is added.

 Enabling corrections by addition brings about a substantial difficulty as the universe of rules that can be potentially added to the
 program can be easily unwieldy or even infinite.
 Hence, we assume that the user (the programmer) provides us with a set of rules
 $\cR$ that can be removed from the program and a set of rules $\cA$ that can be added to the
 program.

\begin{definition}[$(\cA,\cR)$-correction]\label{def:mcs}
 Let $\cA$ and $\cR$ be sets of rules and $P$ be an inconsistent logic program.
 An \emph{$(\cA,\cR)$-correction} of $P$ is a pair $(M_r,M_a)$
 such that $M_r\subseteq \cR$ and $M_a\subseteq \cA$ and the program
 $(P\setminus M_r)\cup M_a$ is consistent.

 An $(\cA,\cR)$-correction $(M_r,M_a)$ is \emph{minimal} if for any $(\cA,\cR)$-correction
 $(M'_r,M'_a)$ such that $M_r'\subseteq M_r$ and $M'_a\subseteq M_a$,
 it holds that $M_a=M'_a$ and $M_r=M'_r$.
\end{definition}

 We refer to the problem of calculating a minimal correction
 as \problem; it accepts as input
 sets $\cA$, $\cR$, and a program $P$ and outputs a pair $(M_r,M_a)$
 in accord with \autoref{def:mcs}.

We show how to translate \problem to maximal consistent subset calculation.
Construct a program $P'$ from $P$ as follows\footnote{Similar construction appears elsewhere~\cite{syrjanen2006debugging}.}.
\begin{enumerate}
  \item Introduce fresh atoms~$s_r^\rr$ for each $r\in\cR$ and $s_r^\ra$ for each $r\in\cA$.
  \item Replace each rule $r\in\cR$ with  $\head(r)\leftarrow s_r^\rr,\ \body(r)$
  \item Replace each rule $r\in\cA$ with  $\head(r)\leftarrow \aspnot s_r^\ra,\ \body(r)$.
\end{enumerate}

Let $\cS=\comprehension{s_r^\rr}{r\in\cR}\union\comprehension{s_r^\ra}{r\in\cA}$.
Then any  maximal consistent subset $\cL$ of~$\cS$ w.r.t.\ $P'$ gives
us a solution to \problem by setting
$M_r=\comprehension{r}{s_r^\rr\notin\cL}$ and $M_a=\comprehension{r}{s_r^\ra\notin\cL}$.
It is easy to see why that is the case. From the definition of maximal
consistent subset, there must be an answer set $\mu$ of the program
$P'\union\comprehension{s.}{s\in\cL}$. If $s_r^\rr\in\mu$ for $r\in\cR$, then
$\mu$ also satisfies the original rule~$r$, hence it is not necessary to remove
$r$ to achieve consistency. Similarly, if $s_r^\ra\in\mu$, then the body
of $\head(r)\leftarrow\aspnot s_r^\ra,\ \body(r)$ is false, i.e.\ the rule is
ineffective, and therefore it is not necessary to add the rule $r$ to the
original program to achieve consistency. Conversely, adding any further $s_r^\rr$
or $s_r^\ra$ to $\mu$ leads to inconsistency due to the definition of maximal
consistent subset. Hence, $M_r$ and $M_a$ defined as above are irreducible.

\subsection{Minimal Corrections for Program Inputs}
Here we come back to the problem that we used to motivate the paper in
introduction, which is to calculate corrections of program's data  so that the
program becomes consistent. Hence, we assume that there is a program $P_e$---
representing the encoding of the problem and a program $P_d$---representing data
to the encoding, consisting only of facts. Assuming that $P_e\union P_d$ is
inconsistent, we wish to identify correction to $P_d$ that would make it
consistent.

For such, we use the concept of minimal corrections as described above.
What is needed is to specify the sets $\cR$ and $\cA$.
To specify the set $\cR$, the user just needs to identify facts in the given input that can be removed.
This can for instance be done by giving predicate names.
Specifying the set $\cA$ is more challenging because it may comprise facts that
do \emph{not} appear in the program.
Hence, we let the user give expressions~$E$ of the form $p(A_1):t(A_2)$
where~$p$ and~$t$ are predicate symbols and $A_1$,$A_2$ appropriate arguments.
The set $\cA$ is obtained by an additional call to an ASP solver.
This call constructs a program $P'$ just as above
and add the rule
$
        \atleast\left(
         \comprehension{s_r^\rr}{r\in\cR}
         \cup
         \{ E_1,\dots,E_u \}
        \right)
        $.
If this call is successful,
the obtained answer set contains facts of the form $p(A_1)$ that constitute the set $\cA$.

%% file: experiments.tex
\section{Experimental Evaluation}\label{sec:experiments}
The proposed algorithms were implemented using the state-of-the-art ASP solver
\textsf{clingo}, version \textsf{4.2.1}. Algorithms~\ref{alg:a}--\ref{alg:p} were implemented
as presented, i.e.\ the ASP solver is invoked repeatedly from a \textsf{python}
script. Computing a cardinality-maximal consistent subset was done by a single
call to \textsf{clingo}, using the minimization rule syntax ({$:\sim$}).
For readability we refer to the individual approaches by single letters,
\autoref{alg:a}--letter \textbf{a} (\textbf{a}tleast-1 constraint);
\autoref{alg:u}--letter \textbf{u} (\textbf{u}nit addition);
\autoref{alg:p}--letter \textbf{p} (\textbf{p}rogression);
the ma{\bf x}imization approach is denoted by \textbf{x}.
The evaluation was carried out on machines with  Intel Xeon 5160 3GHz and 4GB of memory.
The time limit was set to 1000 seconds and the memory limit to 2GB.
The experimental evaluation considers several problems from the 2013 ASP
Competition\footnote{\url{https://www.mat.unical.it/aspcomp2013/FrontPage}}.
The following classes were considered.

\emph{Solitaire.}
In the game a single player
moves and removes stones from the board according to the rule that a stone
can be moved by two squares if that move jumps over another stone; the stone
being jumped over is removed (similarly to \emph{Checkers}).
The problem to be solved in the context of this game is to perform $T$ steps
given $T$ and an initial configuration. For instance, the problem does
\emph{not} have a solution if the board contains only one stone and at least one move
is required to be made.

The size of the board is 7x7 with 2x2-size corners removed, thus
comprising 33 squares. The initial configuration of the board is given by the
predicates $\emptyPred(L)$ and $\fullPred(L)$, where $L$ is a square of the board.
The set $\cR $ was specified as containing all the input's
facts using $\emptyPred$ and $\fullPred$.
The set $\cA$ was specified by the expressions
$\emptyPred(L)\,:\,\locationPred(L)$ and
$\fullPred(L)\,:\,\locationPred(L)$.

To generate unsatisfiable instances of the game, initial board configurations
were generated randomly and  the parameter~$T$ (number of steps) were fixed.
We considered $T=12$ and $T=16$; $100$ instances were generated for each~$T$ and
only inconsistent were considered. This process was repeated for other
benchmarks.

\emph{Knight tour with holes.}
The input is a chessboard of size~$N\times N$ with $H$~holes.
Following the standard chess rules, a knight chess piece must
visit all non-hole position of the boards exactly once and return to the initial
position (which may be chosen). The objective of correction was to remove as few holes as possible so that the tour
is possible. Random hole positions were generated for fixed $N$ and $H$.
Instances with $N=8$, $H=4$ and $N=8$, $H=10$ were considered.

\emph{Graceful graphs.}
Given a graph~$(V,E)$ the task is to determine whether it is possible to label its
vertices with distinct integers in the range $0..|E|$ so that each edge is
labeled with the absolute distinct between the labels of its vertices
and sot that all edge labels are distinct (such graph is called \emph{graceful}).
The correction, for a given graph that is not graceful, tried to find its subgraph that
maintains as many edges as possible (a single-edge graph is graceful).
Considered instances were generated with $|V|=10$ and $|E|=50$
and $|V|=30$ and $|E|=20$.

\emph{Permutation Pattern Matching.}
The problem's input are two permutations $T$ and $P$. The task is to determine
whether $T$ contains a subsequence that is order-isomorphic to~$P$.
For the purpose of correcting we consider $T$ and $P$ so that $P$ is
not an order-isomorphic subsequence of $T$ with the objective of adding and
removing elements to $P$ so that $P$ satisfies the condition.
Considered instances were generated with $|T|=16$ and $|P|=10$
and $|T|=20$ and $|P|=15$.

\begin{table}[t]
\centering
\input{table}
\vspace{5pt}
\caption{Number of solved instances for each instance class.
  The parameters of the class are given in square brackets while total number of the
  instances in a clause are in parentheses.
}\label{tab:all}
\end{table}

\begin{figure}[t]
\centering
  \includegraphics[width=.79\textwidth]{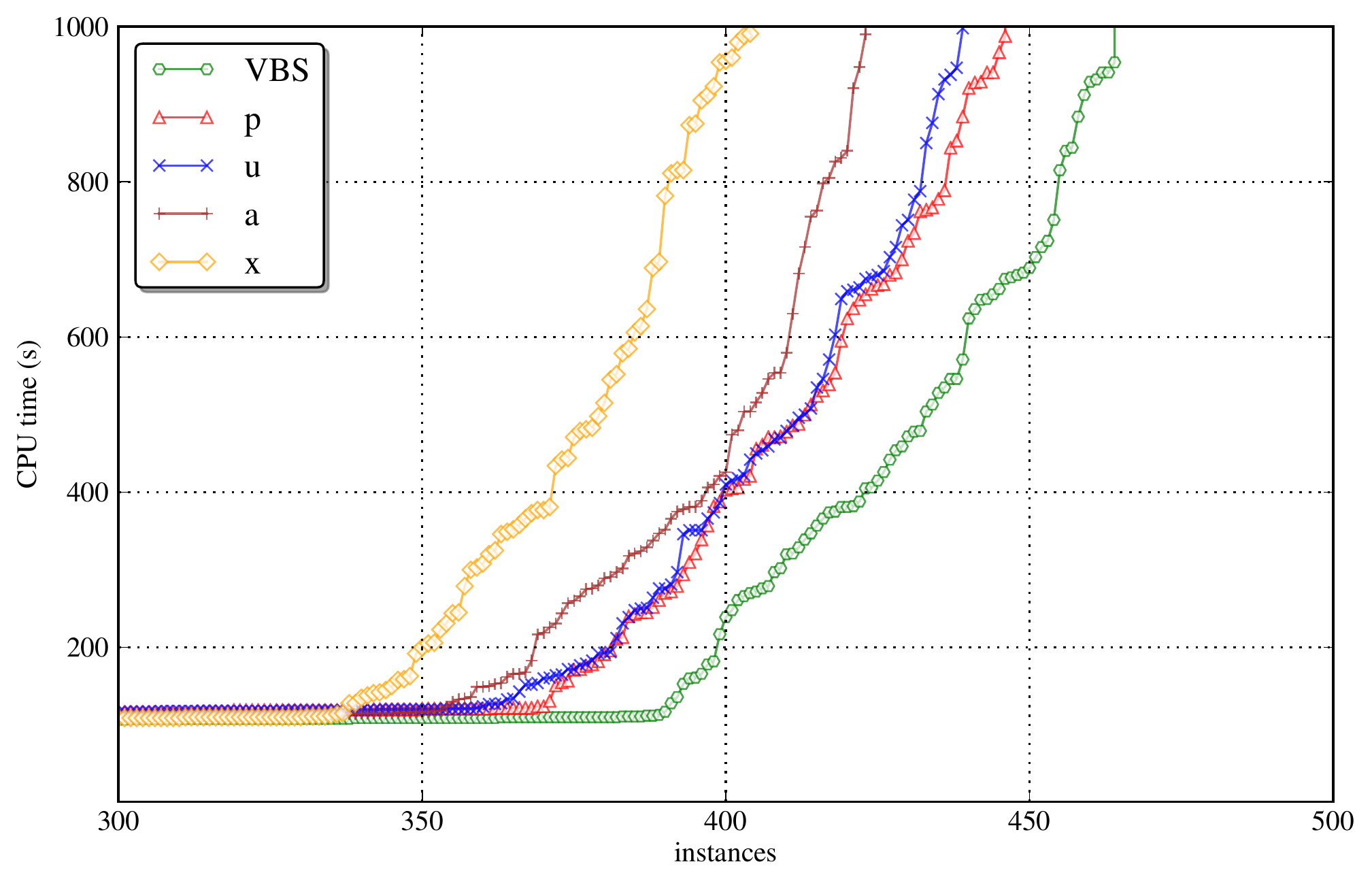}
  \caption{Cactus plot of the overall results where the first 300 easiest
    instances were cut off.
  A point in at coordinates $(n,t)$ means that there are $n$ instances
such that each is solved in time less than $t$.}\label{figure:cactus}
\end{figure}

\begin{figure}[t]
\centering
  \includegraphics[width=.79\textwidth]{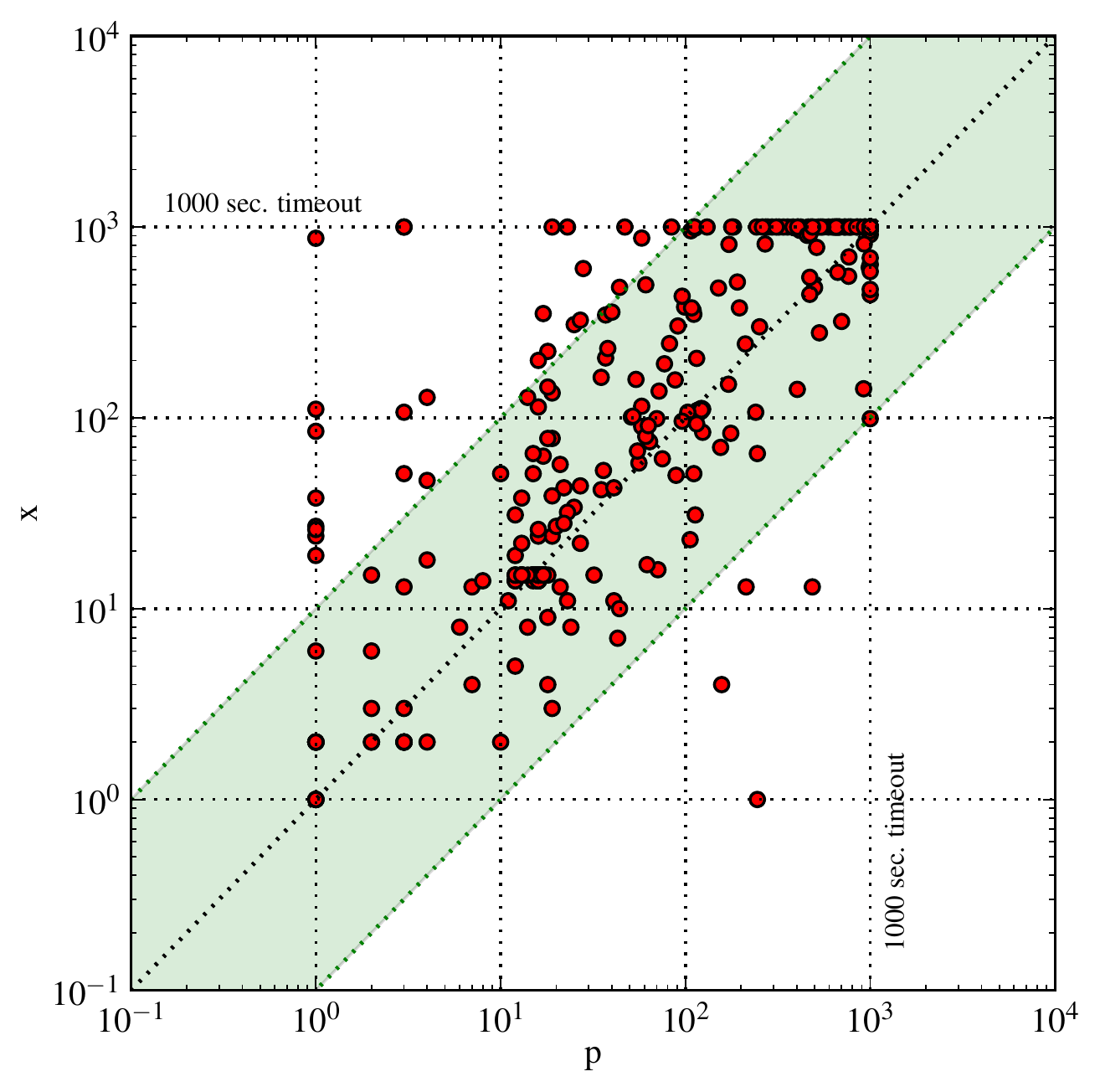}
  \caption{Scatter plot comparing approaches x and p (logarithmic scale).
           }\label{figure:scatter}
\end{figure}

A more detailed overview of the results can be found on authors' website\footnote{\url{http://sat.inesc-id.pt/~mikolas/jelia14}}.
\autoref{tab:all} shows the number of solved instances by each of the approaches
for the different benchmark families. The last column shows the \emph{virtual best solver (VBS)},
which is calculated by picking the best solver for each of the instances.
While the progression-based algorithm ({\it p}) is  clearly in the lead, it is not a winner for all the instances (or even classes of instances).
Indeed, the virtual best solver enables us to solve almost 20 more instances compared to the progression-based approach.
The strength of progression  shows on benchmarks with larger target sets.
Such is the case for the class \textsf{graceful
graphs [10,50]}, where the target set is the set of edges~$E$ with $|E|=50$.
On the other hand, the approaches {\it a} and {\it u} do not aim at minimizing the number of
calls to the ASP solver but represent calls that are likely to be easier.
In {\it u}, the solver only must make sure that one fact from the target set is set to
true. In {\it a}, the solver can in fact chose which fact should be set to true.
These algorithms are the two best ones for \textsf{solitaire~[16]}.
\autoref{figure:cactus} shows a cactus plot across all the considered families (300 easiest instances were cut off for better readability).
This plot confirms data in \autoref{tab:all}. Progression provides the most
robust solution but at the same time, there is a significant gap between
progression and the virtual best solver.

Out of the considered algorithms, cardinality maximization (x) performs the
worst. \autoref{figure:scatter} compares progression to cardinality maximization
in a scatterplot. There are some instances where maximization performs well but
overall progression dominates maximization; by orders of magnitude in a number of cases.

\begin{table}[t]
\centering
\input{distrib}
\vspace{5pt}
\caption{Distribution of how difference between $|M_a|+|M_r|$ for a--p
        and for x.}\label{tab:distrib}
\end{table}

While Algorithms~\ref{alg:a}--\ref{alg:p} give us an advantage over the
cardinality maximization approach, the data so far does not show how the minimal corrections
differ from the minimum-cardinality corrections in size. This is shown
in \autoref {tab:distrib}. The table shows a distribution of how much the value
$|M_a|+|M_r|$ differs from the value obtained from the approach x. For instance,
row 3, column p shows that in 12 cases the value differs by 3 from the minimum
one. Naturally, these data only come from instances where x and the given approach
finished with success. The distributions have a characteristic ``heavy tail''.
In the majority of cases , the actual minimum is
obtained (over 300 for all the approaches). Outliers exists but they are small in numbers.

%% file: table.tex
\begin{tabular}{|l||c|c|c|c||c|}
\hline
\textbf{Family}
& {\bf a} & {\bf p} & {\bf u} & {\bf x} & {\bf VBS}\\\hline
\hline
knight [8,10] (95)
&
74
&
75
&
{\bf 78}
&
60
&
80
\\
\hline
knight [8,4] (51)
&
7
&
{\bf 13}
&
{\bf 13}
&
7
&
14
\\
\hline
patterns [16,10] (100)
&
{\bf 100}
&
{\bf 100}
&
{\bf 100}
&
{\bf 100}
&
100
\\
\hline
patterns [20,15] (100)
&
{\bf 100}
&
{\bf 100}
&
{\bf 100}
&
{\bf 100}
&
100
\\
\hline
solitaire [12] (18)
&
{\bf 18}
&
{\bf 18}
&
{\bf 18}
&
17
&
18
\\
\hline
solitaire [14] (16)
&
{\bf 12}
&
9
&
11
&
4
&
13
\\
\hline
graceful graphs [10,50] (100)
&
57
&
{\bf 75}
&
63
&
62
&
83
\\
\hline
graceful graphs [30,20] (57)
&
56
&
{\bf 57}
&
{\bf 57}
&
55
&
57
\\
\hline
\hline
{\it total (537)} & 424 &  {\bf 447} & 440 &  405 &  465\\
\hline
\end{tabular}

%% file: distrib.tex
\begin{tabular}{|l||c|c|c|}
\hline
$\mathbf{\Delta}$ & {\bf a} & {\bf p} & {\bf u}\\\hline\hline 
0 &  300 & 328 & 332\\\hline
1 &  19  & 16  & 20\\\hline
2 &  38  & 39  & 30\\\hline
3 &  11  & 12  & 14\\\hline
4 &  21  & 16  & 15\\\hline
5 &  8   & 7   & 7\\\hline
6 &  17  & 9   & 8\\\hline
7 &  1   & 1   & 1\\\hline
8--20 &   9   & 19  & 13\\\hline
\end{tabular}

%% file: related.tex
\section{Related Work}\label{sec:related}
The proposed approach is closely related to the work on debugging on answer set
programs~\cite{DBLP:conf/asp/BrainV05%
,DBLP:conf/iclp/Brain06%
,DBLP:conf/aaai/GebserPST08%
,syrjanen2006debugging%
,DBLP:journals/tplp/OetschPT10%
,DBLP:journals/tplp/BrummayerJ10}.
Namely, the approach of Brain~et~al., which also may produce new facts in a
correction~\cite{DBLP:conf/iclp/Brain06}.
Such facts, however, may only be heads of existing rules. Hence, our approach to
data debugging by corrections gives the user a tighter control over the given set of inputs.

Nevertheless, the existing work on ASP
debugging (in
particular~\cite{syrjanen2006debugging,DBLP:conf/iclp/Brain06,DBLP:conf/aaai/GebserPST08}),
also needs to deal with removing redundancy.
For such, the mentioned works use the cardinality-maximal/minimal sets.
Hence, our approach to maximal subset consistency could be applied instead.
Our experimental evaluation suggests that this could improve efficiency of those
approaches.

Other approaches to maximality in ASP exist. In particular, Gebset~et~al.~\cite{DBLP:journals/tplp/GebserKS11} use
meta-modeling techniques to  optimize given
criteria and Nieves and Osorio propose
calculation of maximal models using ASP~\cite{DBLP:conf/lanmr/NievesO07}.
Both approaches, however, hinge on \emph{disjunctive logic programming (DLP)}.
Hence, these approaches require solving a problem in the second level of
polynomial hierarchy. Intuitively, this means worst-case exponential  calls to
an NP oracle. In contrast, all our approaches require \emph{polynomial} calls to an NP
oracle.


Some of the algorithms described in the paper are inspired by work on
computing MSS/MCSes of propositional formulas in conjunctive normal form,
which in term can be traced back to Reiter's seminal work on
model-based diagnosis~\cite{reiter-aij87}.
The well-known \textsf{grow} procedure is described for example
in~\cite{DBLP:conf/padl/BaileyS05} and more recently
in~\cite{biere-vamos12}. However, it is used in many other settings,
including backbone computation~\cite{msjl-ecai10} or even prime
implicant computation~\cite{leberre-fmcad13}.
There has been recent renewed interest in the computation of
MCSes~\cite{mshjpb-ijcai13,msjb-cav13}. Additional algorithms include
the well-known QuickXplain
algorithm~\cite{junker-aaai04,felfernig-aiedam12} or dichotomic
search~\cite{sais-ecai06}.
Clearly, and similarly to the case of ASP, MaxSAT can be used for
computing and enumerating MCSes~\cite{liffiton-jar08}.

Although the paper considers a restricted sample of algorithms for
computing MCS algorithms, adapted
from~\cite{DBLP:conf/padl/BaileyS05,mshjpb-ijcai13,msjb-cav13}, other
options would include the additional approaches summarized above.

%% file: conc.tex
\section{Summary and Future Work}\label{sec:conc}
Motivated by debugging of data to ASP programs, the paper studies the problem
of minimal (irreducible) corrections of inconsistent programs and the problem of maximal
consistency. The two problems are closely related. Indeed, a minimal correction
to an inconsistent program can be calculated by computing a maximal consistency
with respect to only slightly modified program.

We show that algorithms for calculating maximally satisfiable sets (MSSes)
in propositional logic can be adapted to ASP. This is an interesting result as
unlike propositional logic, ASP is not monotone. For three MSS algorithms we show
how they are ported to ASP (Algorithms~\ref{alg:a}--\ref{alg:p}). A similar approach, however, could be used for
other MSS algorithms.

The algorithms for maximal consistency let us then calculate minimal
corrections. For these corrections we assume a general scheme where rules may be added
or removed in order to restore consistency. For evaluation we return to the
initial motivation of the paper and that is the calculation of corrections to data that
make a program inconsistent. A number of instances from various problem classes
were considered. The progression-based algorithm (\autoref{alg:p}) turned out to be the most
effective overall. Nevertheless, it was not a winner for each of the considered classes.
In contrast, the maximum-cardinality approach clearly performed the worst.
Here, however, we should point out that we used the default implementation of minimization
in \textsf{clingo} and it is the subject of future work to evaluate other algorithms.
(A number of MaxSAT algorithms are being adapted to Max-ASP~\cite{DBLP:conf/lpnmr/OikarinenJ09,DBLP:conf/iclp/AndresKMS12}).
Overall, the evaluation suggests that a portfolio comprising the different algorithms
would provide the best solution.

The paper opens a number of avenues for future work.  Irredundancy is needed in
other approaches to
debugging, e.g.~\cite{syrjanen2006debugging,DBLP:conf/iclp/Brain06,DBLP:conf/aaai/GebserPST08}.
It is the subject of future work to evaluate the proposed algorithms also in
these contexts.  The prototype used for the evaluation uses the ASP solver in a
black-box fashion.  It is likely that integrating the algorithms directly into
an ASP solver would give further performance boost.  Similarly, can the
proposed algorithms be made more efficient if the algorithms had an access to
the workings of the ASP solver?

%

 %